# Transitions and Probes in Turbulent Helium

Virginie Emsellem[1], Leo P. Kadanoff[2,†], Detlef Lohse[2,3,⋆], Patrick Tabeling[1], Jane Wang[2,‡]

[1] *Laboratoire de Physique Statistique, Ecole Normale Supérieure, 75231 Paris cedex 05, France*
[2] *The James Franck Institute, The University of Chicago, Chicago, IL 60637, USA*
[3] *Fachbereich Physik der Universität Marburg, Renthof 6, 35032 Marburg, Germany*

Previous analysis of a Paris turbulence experiment [1,2] shows a transition at the Taylor Reynolds number $Re_\lambda \approx 700$. Here correlation function data is analyzed which gives further evidence for this transition. It is seen in both the power spectrum and in structure function measurements. Two possible explanations may be offered for this observed transition: that it is intrinsic to the turbulence flow in this closed box experiment or that it is an effect of a change in the flow around the anemometer. We particularly examine a pair of "probe effects". The first is a thermal boundary layer which does exist about the probe and does limit the probe response, particularly at high frequencies. Arguments based on simulations of the response and upon observations of dissipation suggests that this effect is only crucial beyond $Re_\lambda \approx 2000$. The second effect is produced by vortex shedding behind the probe. This has been seen to produce a large modification in some of the power spectra for large $Re_\lambda$. It might also complicate the interpretation of the experimental results. However, there seems to be a remaining range of data for $Re_\lambda < 1300$ uncomplicated by these effects, and which are thus suggestive of an intrinsic transition.

## I. INTRODUCTION

Recent velocity measurements in highly turbulent helium gas flow by Tabeling et al. ("the Paris group") [1–5] reveal a transition in the turbulent behavior. Their results show that beyond some crossover Taylor Reynolds number $Re_\lambda \approx 700$ the flatness of the velocity derivatives ceases to increase. Such a transitional behavior has never been reported in open flows [6,7], while earlier results on closed flows, also in helium [8–10], have also revealed the existence of a transition at large Reynolds numbers. For the Paris experiment, it has been argued [5,11] that the crossover signals the instability of vortex tubes (worms) [5]; one may speculate also whether it signals the onset of K41 turbulence [12].

This paper is devoted to a further analysis of this crossover, including two main questions: 1. Can the transition be seen in the further analysis of correlation functions? 2. Can it possibly be understood as some kind of effect of the probe?

To discuss the former issue we analyze structure functions $D_n(r) = \langle (u(x+r) - u(x))^n \rangle$ of order n=2,4,6 and the energy spectrum. We try to superpose structure functions and spectra for different Reynolds numbers $Re$. Indeed, the superpositions show that there are *different* forms of the curves above and below the transition, but within each subregion the forms of the curves seem quite similar. The shifts necessary to superpose such curves give a measurement of characteristic length or frequency scales in each measurement. These scales, and their dependence upon Reynolds number, are important because they can help to give insight into the physical origin of the transition.

The central point of this paper is to examine the possibility that the apparent transition might be somehow a reflection of measurement imperfections in the probe. We do this by studying two different mechanisms for probe effects in some detail.

There are two likely sources of difficulties in the measurement, one being that the probe might have insufficient time resolution, the other being that it might be too large. To study the first possibility, we follow the analysis of Grossmann and Lohse [14], who showed the possibility of loss of probe sensitivity in temperature measurements for Rayleigh Benard flow [15,8]. This probe effect occurs as a result of delays and averaging related to thermal diffusion through the partially stagnant gas about the probe. In the Paris experiment considered in the present paper, the probe is a hot wire anemometer, which works by heating the gas around it. Once again, the probe measures temperature changes and can possibly have poor temporal response. This is the first probe effect we analyze in this paper. The second kind of effect we consider involves what happens when the probe size and the dissipative scale becomes comparable. One effect which can arise is vortex shedding behind the probe. In "ordinary" hot wire anemometry (i.e. cylindrical probes working in air), the operating conditions are such that, even at large (overall) Reynolds numbers, there is no vortex shedding behind the sensors [16]. In the Paris experiment, the Reynolds number based upon probe size is large enough so that it is reasonable to ask about vortex shedding. This issue has been partially addressed in the past [2,4]; the general conclusion is that if the vortex shedding frequency is outside the turbulent spectrum, no big perturbation is expected. In the opposite case, the vortex shedding mode may couple with the turbulent fluctuations, and perturb the measurement. The vortex shedding is only one of several things which can happen when the probe has a size comparable with the dissipative scale.

In outline then, in the next section, we show evidence for a change in behavior at high Reynolds numbers. The following section is devoted to looking at probe effects which might partially explain the observed change in behavior. In the section 4, additional evidence is drawn from the measured Reynolds number dependence of characteristic scales. In section 5 we demonstrate by a simple estimate that ISR quantities as scaling corrections or PDFs of velocity differences are hardly effected by the transition. The last section is devoted to conclusions.

## II. EVIDENCE FOR TRANSITION

### A. The Flatness

The Paris turbulent velocity measurements are done in low temperature helium gas, following the idea of the Chicago convection experiment [17,15,8] which were built followed upon the experiments of Threifall [18]. The flow is driven by two counterrotating disks of radius $R$. Two different cells are used: Cell 1 with $R = 3.2$cm [3,1,2] and cell 2 with $R = 10$cm [1,2]. The velocity anemometer ("probe") of size $d = 7\mu$m is placed far enough from the boundary layers. The Reynolds number is defined as $Re = \Omega R^2/\nu$. The angular velocity $\Omega$ of the disks, which is about 1-10Hz, remains about the same for all measurements whereas $Re$ is varied by changing the helium pressure and thus the viscosity $\nu$.

The transition was first observed [2] in the properties of the flatness $F$ and the skewness of the velocity derivative. (We focus upon the flatness since the transitional phenomenon was less visible in the skewness.). Fig 1 shows a series of flatness measurements for the small and large cells, plotted as a function of $Re_\lambda$. This Reynolds number is experimentally defined in terms of the RMS fluctuations in the velocity. [1]. Fig 1 incorporates unpublished data recently obtained with the smaller cell. A rough estimate for the experimental uncertainty in flatness is $\pm 15\%$. At lower Reynolds number, the flatness $F$ increases with increasing $Re_\lambda$. Then at a transition value of the Reynolds number, it seems to reach a peak and then level off at higher $Re_\lambda$. The peak flatness is located at $Re_\lambda$ comprised between 550 and 750 for the small and large cells (see fig.1 and also fig. 5 of [2]). These peak values are identified as the $Re_\lambda$ number for some kind of crossover or transition.

The nature of the transition can be seen by studying its characteristic length scale. In particular, the small scale length called $l_{inf}$ (defined in reference [1]) describes the boundary between the ISR and VSR as observed in the third order structure factor. The classical expectation is that a length like this will decrease as $Re^{-3/4}$. This length shows the expected behavior below a characteristic Reynolds number and instead it seems to saturation at higher Reynolds numbers. (See figure 5 in reference [1].) It was also found (see [1]) that the scaling range of apparent ISR behavior seems to saturate and maybe even shrink beyond the transition Reynolds number.

### B. Dissipation spectra and structure functions

The previous work mostly focused on the scalings of the viscous dissipation. Here we seek further evidence of the transition and its nature. We first analyze the energy dissipation spectra $I(k) = k^2 E(k)$ as a function of Reynolds number. (Here, $E(k)$ is the usual power spectrum, obtained as a function of frequency and then translated into a dependence upon wave vectors through its mean velocity following the Taylor hypothesis.) We find that the curves for the dissipation spectra fall into two groups: There is one shape which holds for all curves below the transition and another shape which governs the curves above the transition.

The spectra are calculated over 16 million points, corresponding to 100-1000 large eddy turnover times for each graph. The data is then divided into about $2^9$ sequencies of $2^{15}$ points. The power spectra are computed for each sequence and averaged over different segments. A Hanning window is used in computing the spectra [19]. To collapse the spectra, we determine the peak of the dissipation $I_p$ and its corresponding frequency $k_p$ following the approach of reference [20]. We then rescale the wavenumber and the dissipation spectrum by $k_p$, and $I_p$, respectively. The rescaled spectra are then plotted in Figure 2. The Figure 2a shows the collapse for different

Reynolds number below the transition, whereas the Figure 2b for those above the transition. Figure 2c shows a comparison of the shapes for the two scaling curves. The solid line corresponds to a general fitting form taken from [20,13]:

$$I(k) = I(k_p) \left[ \left(\frac{k}{k_p}\right)^{-5/3} + \alpha \left(\frac{k}{k_p}\right)^{-\beta} \right] \left(\frac{k}{k_p}\right)^2 e^{-\mu k/k_p} \quad (1)$$

where the fitting parameters are $\alpha = 0.7, \beta = -1$, and $\mu$ is determined to make the peak occur at $k/k_p = 1$. In this fit, the second summand reflects the bottleneck energy pileup at the borderline between inertial subrange and viscous subrange [21]. Note the good agreement between the fitting formula and the data holds above the transition, but not below. Figure 2 then shows a situation in which the dissipation spectra changes at the transitional Reynolds number, but has a rather constant form above and below.

The relative shifts in the logarithmic scales define unambiguously the length and energy scale in the system. In a section 4, we shall study the Reynolds number dependence of the length scales in comparison with different theories of what might cause the apparent transition.

To further test the existence of two groups of scaling functions, we examine different structure functions $D_p(r) = \langle (u(x+r) - u(x))^p \rangle$ in the same spirit. First we look at $D_2$. Since $D_2$ and the power spectra contain the same information, the results on $D_2$ should provide us a double check on the previous finding. To collapse the different curves of $D_2$ versus $r$, we use the same method as applied to the dissipation spectra. We find a characteristic distance for each plot by finding the point of the log-log plot at which the slope is a predetermined constant. The constant is picked so that the slope stands at a place in which the viscous dissipation is beginning to occur. We pick the slope equal to 1.5. (Note that the slopes in $D_2$ vary from 2 to $\sim 0.7$.) We then shift the plots on the log scales by the coordinates of this fitting point and overlay the plots. The result of this analysis is shown in figure 3. Once again, we find the data separate into two groups: one for $Re_\lambda < 800$, and another for $Re_\lambda > 1500$. The relative shifts in horizontal and vertical directions define the scale in length and velocity. We return to the length scale in section 4.

We apply the same method to study $D_4$ and $D_6$. We again collapse by fitting that at specific values of the slope in the log-log plot. In these cases, the chosen slopes are respectively 1.5 and 2.5. In figure 4, we see clear separation of two groups. And once again, the separation occurs between $Re_\lambda = 800$ and $Re_\lambda = 1500$. However the figure 5 for $D_6$ do not show such separations. The failure of seeing two groups in $D_6$ somewhat weakens the argument for a simple transition.

Thus we have shown some additional evidence for the existence of a transition in the Paris experiment. But only $D_2$, its Fourier Transform $E(k)$, and $D_4$ indicate a simple transition. The other correlation functions suggest a broadened transition.

It is clear that there is some kind of change or transition centered at $Re_\lambda$ of about 700. The question is, How does one explain the observed transition? Can it possibly be an effect of some behavior localized about the probe? We turn to those issues in the next section.

## III. PROBE EFFECTS

In this section, we shall obtain a variety of order of magnitude estimates. To make these estimates, we will have to compare a characteristic dissipative frequency in the system to the characteristic inverse times produced by the probe itself. We estimate the dissipative frequency as $\omega_d = U/(10\eta)$ in agreement with Zocchi et al.'s spectral measurements. This frequency is then the inverse of the time it takes for a disturbance of size $10\eta$ to move past the probe. Here, we use classically expected [22] relation between the Kolmogorov length $\eta$ and the dissipation rate $\epsilon$. The definition of the length is $\eta = \nu^{3/4}/\epsilon^{1/4}$ with $\eta$ having the approximate value

$$\frac{\eta}{R} = 30 Re^{-3/4}, \quad (2)$$

In our last steps of analysis we shall express our results in terms of the Taylor Reynolds number [1]

$$Re_\lambda = 1.57 Re^{1/2}. \quad (3)$$

From eq. (2) we obtain

$$\omega_d = \frac{U}{10\eta} = \frac{\nu}{R^2} \frac{1}{300} Re^{7/4}. \quad (4)$$

Here and below, we do our estimates by setting the Prandtl number equal to unity.

### A. The thermal boundary layer

Note that the results on $l_{inf}$ resembles those deduced from the temperature measurements in highly turbulent Rayleigh Bénard flow [8]: Beyond a critical Rayleigh number $Ra \approx 10^{11}$, the scaling range of the temperature power spectrum becomes smaller for increasing $Ra$ and the measured dissipative power shows a weaker $Ra$ increase as below the transition. Grossmann and Lohse [14] suggested that this apparent transition might really be an effect caused by the probe used to measure temperatures. The thickness $\delta$ of the boundary layer around the probe sets a diffusive time scale

$$\omega_\delta^{-1} = \delta^2/\kappa \qquad (5)$$

where $\kappa$ is the heat diffusivity. Beyond $\omega_\delta$, the measured spectral strength will be reduced. If $\omega_\delta$ is smaller than the UV spectral cutoff $\omega_d$, this will affect the UV side of the inertial range in between $\omega_\delta$ and $\omega_d$. This explanation of the observed high Rayleigh number data as a possible probe effect is still unproven, but it has certainly never been ruled out.

One might wonder whether a similar effect would affect the results of the Paris experiment. This speculation might be enhanced because simplest estimate for the value of $Re_\lambda$ at the transition of the Chicago experiment is $(10^{11})^{1/4} \approx 700$. By some accident (?!) this is the same number as at the observed Paris crossover.

Why is heat diffusion relevant for the velocity measurements? The anemometer is heated by an electrical current. The faster the fluid is passing by the probe, the more power is needed to keep the probe at some constant temperature, which is larger than the temperature of the surrounding. The probe gauge curve power vs. velocity is given by King's law and is experimentally known [3].

What velocities does the probe measure? It can *not* be the velocity directly at the probe as there will be a viscous boundary layer of thickness $\delta$ around it in which the velocity is very small. The heat generated in the probe has to diffusively penetrate this layer. Thus the probe measures the velocity of the helium which is a distance $\delta$ away from the probe. The length scale $\delta$ again sets the diffusive time scale described by equation (5).

Our problem is to now estimate the important value(s) of $\delta$ and then to see whether the $\omega_\delta$ thereby generated provides an important cutoff on the responsiveness of the probe.

This has been carried out in two different ways. One of us (VE) has done numerical simulations of the flow past the probe, assuming a laminar time-independent flow [23]. The calculation must be done numerically since the Reynolds numbers of the probe

$$Re_{probe} = \frac{Ud}{\nu} = Re\frac{d}{R} \qquad (6)$$

is of the order of 40 at the observed transition point. Here, $d = 7\mu m$ is the size of the probe. A steady state is reached in which the flow is constant and in which there is a constant flow of heat away from the probe. The calculated flow is then perturbed with a sudden upstream rise in temperature. The probe responds and the important result is that the probe response time is shorter than any time resolved in the experiment. The conclusion is that one can assume that the thermal response of the probe is perfect.

The next question is: How large is the viscous boundary layer $\delta$? There are, in fact, two answers. If the probe Reynolds number is of order ten to one hundred, then there are regions with a thin viscous boundary layer in which the flow comes very close to the cylinder. According to Blasius theory this boundary layer thickness should be:

$$\delta \propto d/\sqrt{Re_{probe}}. \qquad (7)$$

The flow behind the probe produces a much larger region of stagnant fluid at rest, with a thickness comparable to the cell size. Thus we have also

$$\delta \propto d. \qquad (8)$$

The frequency produced by these lengths via equation (5) will only matter if they are smaller than or of the order of the Kolmogorov cutoff of equation (4). Since these frequencies differ by a factor of the probe Reynolds number they are quite different. The conditions they generate are different also. If the length $d$ matters, then the frequency cutoff will bother us whenever the Reynolds number, $Re$ obeys

$$Re > \left(300\left(\frac{R}{d}\right)^2\right)^{4/7}, \qquad (9)$$

Conversely if the size of the Blasius boundary layer matters, we will get the less stringent condition

$$Re > (300R/d)^{4/3}. \qquad (10)$$

for the Reynolds number at which the thermal boundary layer of the probe becomes important.

These answers are quite different. If the important distance is $d$, we get from equation (9) an estimate of the critical Taylor Reynolds number, $Re_\lambda$, as $\approx 1000$ for the small cell and $\approx 2000$ for the large cell. The experiment measures a crossover $Re_\lambda$ of order 700. These results are perilously close.

On the other hand, if it is appropriate to use the Blasius length then equation (10) gives a much larger critical Reynolds numbers, in fact a critical value of $Re_\lambda$ which is greater than $10^4$ and which can not be realized in the experiment.

So which length should we use? To see the answer notice that the experiment measures the heat flow out of the probe. This heat flow is much larger in regions in which the boundary layer is thinner. Thus, for larger $Re_{probe}$ the important regions are the ones which have the Blasius effect and are thinned by a factor of $(Re_{probe})^{-1/2}$. The simulation [23] fully supports this point of view. Note also that this line of argument does not apply to the Chicago experiment [15,8] and its analysis [14]. In the Chicago case, the probe measures the average temperature in its environment. Regions of thick boundary layer are as important as thin ones. Therefore, relatively sluggish regions can effect the outcome. In contrast, in the Paris experiment, only the most responsive regions matter.

## B. The dissipation rate test

One can construct a direct test of the probe responsiveness. The energy dissipation rate - denoted by $\epsilon$ - can be measured in two different ways [1]. One involves a measurement in the inertial subrange and determines a quantity which we call $\epsilon_{ISR}$. The other is a dissipative measurement and determines $\epsilon_{VISR}$. Both quantities can be expressed in dimensionless form by writing $c$ as the ratio of $\epsilon$ to $U^3/R$ where $U = \Omega R$ is the large scale velocity. In this way one finds the two different (Reynolds number dependent [24]) ratios

$$c_{\epsilon, ISR} = \epsilon_{ISR}\frac{R}{U^3} \quad \text{and} \quad c_{\epsilon, VSR} = \epsilon_{VSR}\frac{R}{U^3} \qquad (11)$$

One can find $\epsilon_{ISR}$ by using the fact that the third order structure function $D^{(3)}(r)$ obeys the Kolmogorov structure equation [22]

$$D^{(3)}(r) = -4\epsilon r/5 \qquad (12)$$

When the Paris group carries out their ISR measurement their results agree with the expected spatial scaling and thus enable them to construct $c_{\epsilon, ISR}$. This is plotted in figure 6. In addition, the energy dissipation rate $\epsilon$ is measured by a method which uses the viscous subrange, $\epsilon_{VSR}$, namely by determining $\epsilon$ from the spectrum $E(k)$ as

$$\epsilon_{VSR} = 15\nu \int_0^\infty dk k^2 E(k). \qquad (13)$$

Here, full isotropization has been assumed. Strictly speaking, $\epsilon_{VSR}$ is only based on $\langle(\partial_1 u_1)^2\rangle$ which is (via Taylor's hypothesis [25]) the only experimentally accessible contribution to the strain tensor $\partial_i u_j$. This determination of $\epsilon$ then gives the other dimensionless constant, $c_{\epsilon, VSR}$, which is also plotted in figure 6. If the probe response was cut off at high frequencies, one would expect $c_{\epsilon, VSR}$ to be substantially smaller than $c_{\epsilon, ISR}$. The data shows no support for that hypothesis. So we must conclude that, within experimental error (which is rather substantial), the high frequency response of the probe is satisfactory.

## C. Vortex shedding behind the probe

In this section, we discuss the effect of finite probe size on spatial resolution. We focus particularly upon the effects of vortex shedding from the probe.

Zocchi et al. [1] have observed a series of peaks at the high end of the power spectra. The exact origin of these peaks is unknown. Possible sources are: vortex shedding, vibrations of probe and its support. In more recent experiments the fiber has been strained at a tensile strength ten times larger than before and the peaks have mostly disappeared. This indicates that some the observed peaks were a vibration of the fiber. Here, we study the vortex shedding to estimate where it will occur and the effect it is likely to have.

The frequencies of vortex shedding can be estimated as:

$$f_v = St\frac{U_0}{d} \qquad (14)$$

where $St$ is the Strouhal number, which is typically 0.2 [27]; $U_0$ is the velocity of an ambient fluid, and $d$ is the diameter of the obstacle. We focus upon the shedding from the probe, which has a diameter $d \sim 7\mu$m.

Two conditions must be satisfied for the vortex shedding to be important. First it must be present. Vortex shedding appears when the probe Reynolds number is above $\sim 40$ [27]. In figure 8 we plot the probe Reynolds number $Re_{probe}$ as a function of $Re_\lambda$. The horizontal dashed line corresponds the known [27] onset value 40. Its crossing with the experimental data gives a reading of transitional value in $Re_\lambda$, which is between 600 and 800 for small cell, and 600-1100 for large cell. These ranges are close to the observed transition in flatness. Next, the shedding frequency must be in an observable range of frequency. by using typical values for $U_0 \sim 1m/s$ in equation (14), we estimate the vortex shedding frequency to be about $15kHz$, which is in the observable range of frequency.

We then ask when does vortex shedding have an important effect upon the flatness? Since the flatness is measured as a scale on the order of typically $5\eta$, we expect the vortex shedding begin to affect the flatness when the vortex shedding frequency (eq 14) is comparable with the Kolmogorov frequency ( eq. 4). Under Taylor's hypothesis, this is equivalent to comparing the shedding wavelengthes $l_v$ and Kolmogorov length $\eta$. The shedding wavelength is given by:

$$l_v = \frac{U_0}{2\pi f_v} \sim 1.2d, \qquad (15)$$

In figure 9 we plot $\eta$ as a function of $Re_\lambda$. Once again, we find that the line $1.2d$ crosses the data around the same range of transitional value in $Re_\lambda$ as seen in the flatness. Thus, the vortex shedding shows promise of explaining the position of the transition.

Next we want to know how does the vortex shedding affect the measurements of the flatness. Let us decompose the velocity signal $U(t)$ as

$$U(t) = V(t) + A\sin(\omega_v t) \qquad (16)$$

where $V(t)$ denotes the velocity in the absence of shedding and the other term might reflect, for example, a shaking of the probe with angular frequency $\omega_v$. We know that the intrinsic signal, $V(t)$ gives a very large flatness, of order ten. On the other hand a sinusoidal signal, like the term in $A$, will give a much smaller flatness. Thus, we should expect that such an additional term would tend to reduce the flatness as seen in the experiments. A careful calculation bears out this point [28].

However the figures 8 and 9 show some puzzling features. Note the probe Reynolds number is defined as in equation (6). For a fixed geometry we expect $Re_\lambda$ to be a fixed constant times $Re^{1/2}$. Thus all the black dots (corresponding to the small cell) in figure 8 should fall on one straight line with slope 1/2 and all the white dots should fall on another, lower by an amount corresponding to the log of the ratio of the cell sizes, $\log 3$. So why do the large cell data fall onto two lines? Figure 1 and 8 collects data measured in two distinct cells : one is 3 cm in radius, and the other one is 10 cm in radius. They cover four series of experiments performed in the period 1994-1995; there is no simple relation between $Re_\lambda$ and $Re$, because, for the same cell size, we may have substantially different integral scales (due to the fact that the rims have not the same size from one series to the other); moreover, we have worked with different velocity fluctuation rates. All this explains why, for a given cell size, we have not a simple relation between $Re_\lambda$ and $Re$. However, if we fit the means of the two lines we get a reasonable value of the separation between the curves. This separation realizes our theoretical expectation that the two intersection points should have $Re_\lambda$ values which differ by a factor of the square root of the size ratio or roughly 1.7.

On the other hand, according to figure 1 there is no obvious difference in the transition point between the larger cell and the smaller. This lack of difference would be expected if the transition were an intrinsic effect; it is not expected if it results from the probe size via having $Re_{probe}$ with a constant value at the transition. Thus, figure 8 seems to explain the position of the transition as a probe effect, but cannot satisfy us on the question of the size dependence. We do note that the uncertainties in the measurements are large, thus a possible size dependence maybe overlooked.

Now we return to the question on the size of the probe relative to $\eta$. The shedding wavelength (eq. 15) is independent of the Reynolds number. On the other hand, the maximum dissipative frequency gives a scale $10\eta$ which decreases with Reynolds number as $Re^{-3/4}$. If $10\eta$ becomes comparable or smaller than $d$, the measurements of the dissipative quantities, such as the length measurements discussed in the next section, will be influenced by the finite probe size. Figure 9 shows that in fact the probe size is considerably smaller than $10\eta$ so that it is likely that the vortex shedding will give such high frequencies that it is unlikely to affect the measured length or even the flatness(The reader will recognize that we are getting onto dangerous ground, since we are trying to distinguish between $\eta$ and $10\eta$ in an order of magnitude argument).

What does one expect to see in the length measurements if probe effect is relevant? As we have already seen, the vortex sheddings introduce to the system an external length scale which is independent of Reynolds number. This length scale becomes observable if the measurements of interest is on the order of the probe size. This suggests a saturation of $l_{inf}$ above the transition, which is consistent with Figure (5) in reference [1]. Similarly, it predicts that the length scales divided by the Kolmogorov scale increase with Reynolds number as $Re_\lambda^{3/2}$. The comparisons will be done in the next section.

Finally we like to remark that a sharper conclusion concerning the vortex shedding (or other finite probe size effects) can be drawn if we compare two experiments with and without vortex shedding at the same $Re_\lambda$. In other words, we want to find out whether the probe Reynolds number becomes a relevant parameter above the transition. One hopes that at $Re_\lambda$ greater than the observed transition, by either varying the cell size or the probe size, we can change the probe Reynolds number such that it is much greater than the critical value 40 for one case and much smaller than the critical value for another. If the transition is independent of the two situations, then we can believe $Re_{probe}$ is not a parameter determining the transition. and thus the transition is very likely to be intrinsic. So far the current available data, as shown in figure 8, do not yet provide us with such an ideal situation. We see that for $Re_\lambda > 600$, $Re_{probe}$ is roughly the same for both systems, thus we are unable to distinguish the two cases. We hope that future experiments will eventually clarify these ambiguities.

## IV. CHARACTERISTIC LENGTHS: MEASUREMENTS VS. PREDICTIONS

In this section, we discuss the the characteristic lengths displayed by the turbulence data. Below the transition, we expect that the Kolmogorov dissipation length provide the characteristic scale for all short-distance phenomena. Beyond the transition, we expect that another length might enter the problem. In both the vortex shedding approach and the thermal boundary layer (TBL) approach, this other length turns out to be of the order of the size of the probe[1]. In this section, we measure all lengths in units of the dissipation length. Therefore, we expect to see a constant value for this ratio below the transition. If the probe dominates the behavior by either of the mechanisms discussed here, we expect the ratio of characteristic length to dissipation length to increase as $Re_\lambda{}^{3/2}$.

---

[1] This result is obtained for the TBL by substituting equation (7) into equation (5) and multiplying by the integral scale velocity. The result is a length which is essentially the probe size.

Our first method defines the dissipation lengths by identifying the maximum energy dissipation scale $k_p$ as we did to collapse the dissipation spectra. Figure 10 shows the inverse of $k_p$ normalized by $k_\eta$ as a function of Reynolds number. The overall behavior can be fit by a constant as shown by the horizontal line. If the probe affect the scales when $Re_\lambda > 700$) we might expect the ratio to slope upward as shown. Either a constant or an upward slope is equally well supported by these data.

Similarly, we can determine other lengths scales by identifying the relative shifts used to collapse $D_2$, $D_4$, and $D_6$ as discussed in section 2. These lengthes normalized by $\eta$ vs. Re are shown in figure 11. Again, a horizontal line is draw across them, and compared with the prediction of the two probe scenarios. These data do not support the notion of an upward slope. Hence they suggest that the scales of order $30\eta$ are not affected by probes unlike the flatness measurements, which correspond to scales of $5\eta$.

Next, we define the cross-over length to be the length $r_m$ corresponding to the minima of the 2nd log-log derivative of $D_2$. Here, $r_{min}$ should capture the cross-over from a $r^2$ scaling behavior at small $r$ to a $r^{0.7}$ scaling in the inertial range. The Reynolds number dependence of $r_{min}$ normalized by $\eta$ is shown in figure 12 in comparison with the probe prediction. No evidence for a probe effect is seen.

The Reynolds number dependence of all these length ratios seems to be consistent with a constant or the probe-predictions within the scatter of data.

Another related length scale $l_{inf}$ defined in reference [1] did show a saturation beyond the transitional Reynolds number, this is a suggestive of the existence of an external scale in the system. The plot of $l_{inf}$ is the figure 5 in ref. [1]

## V. CAN THE TRANSITION BE SEEN IN THE ISR?

The transition in the flatness of $\partial_1 u_1$ at $Re_{\lambda c} = 650$ is rather pronounced, as seen from figure 1. How does this transition towards K41 in a VSR quantity reflect in pure ISR quantities? In this section we will see that for experimentally reachable Reynolds numbers the transition can hardly be expected to be visible in ISR quantities as velocity structure function exponents and PDFs of ISR quantities. This finding is in agreement with the measurements of Tabeling et al. [4,2,1]. Thus the ISR results of the Paris experiment do not contradict the existance of the transition towards K41.

For our estimate here it is sufficient to sketch the flatness $F(r) = \langle v_r^4 \rangle / \langle v_r^2 \rangle^2$ as follows [29]:

$$F(r) = \begin{cases} = F_{sat} = 10 & \text{for} & r \leq 10\eta \\ \sim r^{\zeta_4 - 2\zeta_2} & \text{for} & 10\eta < r < L \\ = F_\infty = 3 & \text{for} & r \geq L \end{cases} \quad (17)$$

$L$ is the integral length scale. From [1] we have $L = 4cm = 0.4R$ for the large cell, independent of $Re_\lambda$. From the sketch (17) we immediately obtain the $Re_\lambda$ dependence of the scaling corrections to K41,

$$\delta(Re_\lambda) = |\zeta_4(Re_\lambda) - 2\zeta_2(Re_\lambda)| = \frac{\lg(F_{sat}/F_\infty)}{\lg(L/10\eta(Re_\lambda))}. \quad (18)$$

The ratio $L/10\eta$ scales like

$$\frac{L}{10\eta} = c Re_\lambda{}^{3/2}. \quad (19)$$

Rather than taking $c$ from eqs. (1) and (2) of section 2 we adopt it to the experimental value of $\delta$ for the Reynolds number $Re_\lambda = 650$ of the transition. From [4] we have $\zeta_4 = 1.25$ and $\zeta_2 = 0.70$. Thus $\delta = 0.15$ and $c = 0.185$. The $Re_\lambda$ dependence of the scaling correction $\delta(Re_\lambda)$ is very weak and approaches its K41 value only logarithmically, $\delta(Re_\lambda) \propto 1/\lg Re_\lambda$, suggesting that $1/\log Re_\lambda$ rather than $1/Re_\lambda$ is the small parameter in turbulence. Some numbers for still experimentally reachable Reynolds numbers are given in table 1. This weak decrease is in agreement with the slight experimental decrease of the scaling correction $\delta \zeta_n$ as shown in figure 8 of [4].

Now we focus on the PDF of $v_r$. In figure 2 of ref. [2] no detectable dependence of the PDF for $v_r$ for *fixed* scale $r = 490\mu m$ on the Reynolds number was noticed. With our above sketch (17) of $F(r)$ we readily calculate

$$F\left(\frac{r}{L}, Re_\lambda\right) = 3 \left(\frac{r}{L}\right)^{-\delta(Re_\lambda)}. \quad (20)$$

The Reynolds number dependence of $F(r/L, Re_\lambda)$ for fixed $r/L = 490\mu m / 4cm = 0.012$ again is very weak, table 1.

Finally, we give as a characteristics of the PDF also the stretching exponent $\beta$ in a parametrization

$$PDF(v_r) \propto \exp\left(-\left|\frac{v_r}{v_r^0}\right|^\beta\right) \quad (21)$$

which is well known to fit the tails of experimental PDFs. $\beta = 2$ means Gaussian PDF. The flatness and the stretching exponents are connected by [29]

$$F = \frac{\Gamma(1/\beta)\Gamma(5/\beta)}{(\Gamma(3/\beta))^2}. \quad (22)$$

The stretching exponent $\beta$ again only very weakly depends on the Reynolds number, table 1. They must be directly compared with figure 2 of ref. [2] where also hardly any dependence is seen.

## VI. CONCLUSIONS

To summarize this section: The Paris transition towards K41 can not be expected to be pronouncedly seen in pure ISR quantities as PDFs for fixed $r$ or scaling exponents $\zeta_n$. To look for a similar transition in other geometries one should thus focus on VSR quantities as e.g. the (hyper)flatnesses.

## VI. CONCLUSIONS

The Paris experiments show a transition in behavior for $Re_\lambda \approx 700$. This transition can be seen in the flatness and indeed in measurements of individual moments like the power spectrum, $D_2$, $D_4$, and $D_6$. However, one might worry that the apparent transition was caused by the finite size or finite frequency resolution of the probe. We have analyzed herein two effects which may affect the measurement performed in the Paris experiment: a. the thermal response of the probe, which imposes a limitation on the temporal resolution and b. the vortex shedding behind the probe, which distorts the spatial resolution of the probe. What should we conclude?

The two probe scenarios discussed in this paper both sound plausible. We cannot prove they are the direct cause of the transition, and in fact, we find partial evidence against each. In the end, we conclude that the measurements up to $Re_\lambda \sim 1300$ are probably not perturbed by the two effects which are discussed in the present paper.

We suspect the thermal boundary layer effect to be relevant at higher Reynolds number $\sim 2000$, but not in the region where the transitional behavior is observed. The key point here is the presence of a boundary layer which enhances the thermal exchanges between the fiber and the surrounding fluid. This is confirmed by numerical analysis [23], which showed that the thermal frequency response is outside the turbulent spectrum in a comfortable range of Reynolds number around the transition. This is further checked by the observation that the dissipation rate seems correctly measured in the same range of Reynolds number; therefore the thermal response of the probe seems appropriate to the measurement of dissipative quantities.

Vortex shedding probably do not affect the measurement in an appreciable range of Reynolds numbers around the transition, either. Although it gives the right transition value in $Re_\lambda$ and produces the observed decrease of flatness with $Re_\lambda$, our basic argument against it is using the fact that the transitional $Re_\lambda$ for small and large cell are roughly the same within the experimental error. On the other hand, this argument is somewhat weakened due to the large error in the experimental measurements. We also remark that a set of improved experiments show less of the anomalous peaks in the spectra, and yet the transition persists.

Perhaps neither the vortex shedding nor the thermal layer is in itself the right explanation. However, the transition does occur when the probe Reynolds number is high (of order 40) and when the dissipation length is of the same order of magnitude as the probe size. There are likely to be many other possible effects, not explored here, which only depend on two essential ingredients: 1) the finite size of the probe, 2) the injection of energy in the small scale comparable to the size of the probe. Thus, another effect of finite probe size or response might intervene and produce a false signal of a transition.

On the other hand it is entirely possible (and likely) that the transition observed is real and has nothing to do with probe effects.

To more fully understand the nature of the transition, we will require further experiments with closed flows. They are likely to involve Helium and probes similar to those employed here. Clearly, it would be very desirable if a major piece of the next experiments were devoted to understanding further the behavior of the probes, and of the flow in their neighborhood, and how this flow evolved with Reynolds number. To fully realize the potential of the experimental method, we need further development of the technique for using probes like these. Perhaps one can probe the velocity and temperature field around large probes with tiny ones to further understand the probe effects.

**Acknowledgements:** Very helpful exchanges with F. Belin, L. Biferale, B. Castaing, J. Eggers, S. Grossmann, K. R. Sreenivasan, G. Stolovitzky, H. Willaime, and G. Zocchi are acknowledged. We also thank the groups in Paris and Chicago for their hospitality during visits from one institution to another. This work has been supported by DOE contract number DE-FG02-92ER25119, by the MRSEC Program of the National Science Foundation under Award Number DMR-9400379, by the DFG through its SFB185, and the NATO grant under award number CRG-950245.

† email: leo@control.uchicago.edu
⋆ e-mail: lohse@cs.uchicago.edu
‡ email: jw@control.uchicago.edu, also, the corresponding author.

**Table 1**
Reynolds number dependence of ISR characteristics for various Reynolds numbers beyond the transition. Only weak dependence is detected.

| $Re_\lambda$ | $\delta(Re_\lambda)$ | $F(490\mu m, Re_\lambda)$ | $\beta(Re_\lambda)$ |
|---|---|---|---|
| 650 | 0.150 | 5.82 | 1.02 |
| 1000 | 0.139 | 5.54 | 1.06 |
| 1500 | 0.130 | 5.32 | 1.09 |
| 3000 | 0.117 | 5.03 | 1.14 |
| 5000 | 0.109 | 4.85 | 1.18 |

FIG. 1. The flatness of the velocity derivatives is plotted against $Re_\lambda$. Results for the small cell are given as the black points and for the large one as the empty circles. This same convention is used in the later figures. Each point shown here is the average of three experiments performed at the same $Re_\lambda$, in the same cell, with the same probe.

FIG. 2. Collapse of dissipation spectra. Parts a and b respectively show overlays of different dissipation spectra for $Re_\lambda$ below and above the transitional value. Part c shows the comparison of two spectra, one from above and one from below.

FIG. 3. Two groups of $D_2$ curves. As in figure 2 this figure contrasts the relatively unvarying behavior above and below the transition with the somewhat larger change which occurs across the transition. In this and next two figures, we use open symbols to indicate the flows below transition, and solids above the transition. (chicago)

FIG. 4. Two groups of $D_4$ curves. As in figure 3 this figure contrasts the behavior within the region above and below the transition with the change which occurs across the transition. (chicago)

FIG. 5. Trying to collapse of $D_6$ curves. As in figure 3 this figure contrasts the behavior within the region above and below the transition with the change which occurs across the transition. This contrast is now constructed for correlation in the velocity cubed. However, we do not see the separation of groups as we did in the previous two figures. (chicago)

FIG. 6. Dimensionless constants which measure the energy dissipation rate, $eps$, found by two different experimental methods. One method uses Integral subrange data, the other using viscous subrange data. The result is plotted as a function of the Taylor Reynolds number. Each point is the average of three experiments performed at the same $rel$, in the same cell, with the same probe. Notice that the two measurement agree within experimental error. Earlier measurements showed an apparent discrepancy in which the dissipative range quantity fell below the integral range one. Those measurements lead to the assumption that probe effects degraded the data at high frequencies [26,14]. These plots show the Paris group's latest measurements performed at smaller fluctuation rates (around 20%, to be compared to 35% for [1]) , and do not reveal any discrepancy between $\epsilon_{ISR}$ and $\epsilon_{VSR}$

FIG. 7. The spectra for $Re_\lambda = 1626$. The arrows marks the positions of the estimated Kolmogorov wavenumber, and the of the vortex shedding.

FIG. 8. The various probe Reynolds numbers we achieve in the experiment for the two cells (small - black points - and large - circles -). The scatter is partly due to the fact that several fluctuation rates are considered in the compilation. The line shows the critical value, $\sim 40$, for the onset of vortex shedding.

FIG. 9. The various ratios $d/\eta$ we achieve in the experiment. Same symbols and same remarks as in Figure 5a.

FIG. 10. The relative shifts used to collapse dissipation spectra divided by $k_\eta$ vs. $Re_\lambda$.

FIG. 11. The relative shifts used to collapse $D_2$, $D_4$ and $D_6$ divided by $\eta$ vs. $Re_\lambda$.

FIG. 12. $r_{min}$ vs. $Re_\lambda$.

[1] G. Zocchi, P. Tabeling, J. Maurer, and H. Willaime, Phys. Rev. E **50**, 3693 (1994).
[2] P. Tabeling *et al.*, Phys. Rev. E53, 1613 (1996).
[3] J. Maurer, P. Tabeling, and G. Zocchi, Europhys. Lett. **26**, 31 (1994).
[4] F. Belin, P. Tabeling, and H. Willaime, Physica D (1996), in press.
[5] F. Belin, P. Tabeling, and H. Willaime, J. de Physique (1996), in press.
[6] C. W. van Atta and R. A. Antonia, Phys. Fluids **23**, 252 (1980).
[7] For a comprehensive recent review see K. R. Sreenivasan and R. A. Antonia, "The phenomenology of small scale turbulence", Ann. Rev. of Fluid Mech., 1996, submitted.
[8] I. Procaccia *et al.*, Phys. Rev. A **44**, 8091 (1991).
[9] X. Z. Wu, L. Kadanoff, A. Libchaber, and M. Sano, Phys. Rev. Lett. **64**, 2140 (1990).
[10] B. Chabaud, A. Naert, J. Peinke, F. Chilla, B. Castaing, B. Hebral, Phys. Rev. Lett. 73, 3227 (1994).
[11] J. Jimenez, A. Wray, P. Saffman, and R. Rogallo, J. Fluid Mech. **255**, 65 (1993).
[12] A. N. Kolmogorov, CR. Acad. Sci. USSR. 30, 299 (1941).
[13] S. Grossmann and D. Lohse, Phys. Fluids **6**, 611 (1994); Phys. Rev. E **50**, 2784 (1994).
[14] S. Grossmann and D. Lohse, Phys. Lett. A **173**, 58 (1993).
[15] B. Castaing *et al.*, J. Fluid Mech. **204**, 1 (1989).
[16] J. O. Hintze, Turbulence, Mc Graw Hill, 1975.
[17] F. Heslot, B. Castaing, and A. Libchaber, Phys. Rev. A **36**, 5870 (1987).
[18] D. C. Threifall, PhD. Thesis, Univ. of Oxford, 1976 (unpublished); J. Fluid. Mech., 67, 1 (1975).
[19] William H. Press, et. al., Numerical Recipes (Cambridge University Press, Cambridge, 1986) page 425.
[20] Z. S. She and E. Jackson, Phys. Fluids A **5**, 1526 (1993).
[21] G. Falkovich, Phys. Fluids **6**, 1411 (1994); D. Lohse and A. Müller-Groeling, Phys. Rev. Lett. **74**, 1747 (1995).
[22] A. S. Monin and A. M. Yaglom, Statistical Fluid Mechanics (The MIT Press, Cambridge, Massachusetts, 1975).
[23] V. Emsellem, C. R. Acad. Sci. Paris, in press (1996).
[24] D. Lohse, Phys. Rev. Lett. **73**, 3223 (1994); S. Grossmann, Phys. Rev. E **51**, 6275 (1995); K. R. Sreenivasan and G. Stolovitzky, Phys. Rev. E 52, 3242
[25] G. I. Taylor, Proc. R. Soc. London A **164**, 476 (1938).
[26] K. R. Sreenivasan (unpublished).
[27] D.J. Tritton Physical Fluid Dynamics (Oxford Scientific Publishers, second edition) page 28.
[28] J. Wang, unpublished
[29] S. Grossmann and D. Lohse, Europhys. Lett. **27**, 347 (1993).